\documentstyle[12pt]{article}

\parskip=\medskipamount                 
\def\7#1#2{\mathop{\null#2}\limits^{#1}}        
\def\ast{\displaystyle *}

\def\beee{\begin{equation}}
\def\eeee{\end{equation}}  
      
\begin{document}

\bibliographystyle{unsrt}                                                    

\begin{center}
{\Large \bf VARIATIONAL PRINCIPLE IN THE ALGEBRA OF ASYMPTOTIC FIELDS}
\footnote{Supported in part by a Semester Research Grant from the General
Research Board of the University of Maryland and by the National Science 
Foundation;
email address, owgreen@physics.umd.edu.}\\
[5mm]
O.W. Greenberg\\
{\it Center for Theoretical Physics\\
Department of Physics \\
University of Maryland\\
College Park, MD~~20742-4111}\\[5mm]
\end{center}

\vspace{2mm}

\begin{abstract}

This paper proposes a variational principle for the solutions of 
quantum field theories in
which the ``trial functions'' are chosen from the algebra of asymptotic fields, 
and illustrates this variational principle in simple cases. 

\end{abstract}

\section{Introduction}

The most-used methods to find approximate solutions of quantum field theories
are based on path integrals\cite{pes,wein,fey,wil,cre}.  
They have many advantages; however the Hilbert 
space and particle structure of field theory are not evident from this point of
view.  Fock space methods, such as the Tamm-Dancoff 
approximation\cite{tamm, dan} and the discretized
light cone quantization approximation (DLCQ)\cite{bpp}
place the Hilbert space and particle structure of the
theory in the forefront; however the covariance of the theory is not evident.
The functional Schr\"odinger picture allows intuitive guesses about the form of
the solutions of a field theory to be incorporated, but this method also fails
to be explicitly covariant.  

Another method, less developed than those just
mentioned, is the expansion in normal-ordered asymptotic fields, the ``Haag
expansion'' or ``N-quantum approximation,'' which applies directly only to
theories without zero-mass particles. In particular, it does not apply 
directly to gauge theories. Nonetheless, this method can be extended to gauge 
theories; indeed it has been applied to quantum electrodynamics.  Using the Haag
expansion, one works in the algebra of asymptotic fields, and can keep creation 
and annihilation parts of operators on the same
footing, since one can choose not to apply elements of the algebra to the vacuum,
where terms with annihilation operators would annihilate the vacuum and be lost.
By contrast, in the
Tamm-Dancoff approximation, the annihilation parts destroy the vacuum and 
disappear from the calculation.
In addition to losing explicit Lorentz invariance, this asymmetric treatment of 
the annihilation and creation parts of the fields destroys crossing symmetry.
The DLCQ method also treats annihilation and creation operators asymmetrically; 
in addition, it suffers from a lack of explicit covariance.  This complicates
renormalization considerably.

Just as variational methods have been used in other approaches to quantum field
theory, this paper proposes a variational principle based on the Haag 
expansion. 
Variational principles in quantum mechanics are powerful ways to go beyond 
perturbation theory.  In quantum mechanics, the solutions lie in a Hilbert 
space of functions and the trial functions are chosen from this space.  For
example, to approximate the ground state, one can choose a wave function $\psi$ 
that depends on some parameters and determine the parameters by minimizing the 
ground state energy\cite{schiff}.  The exact ground state energy will be less 
than the approximate energy at the minimum.  Many attempts have been made to 
carry over this approach to quantum field theories.  In field theory, the 
ground state wave function is replaced by a vacuum wave functional that can
depend on functions as well as parameters.  Minimizing the vacuum energy
determines the functions and parameters and yields approximate information 
about the solution of the theory\cite{tik,yee,kog,kk}.  This approach is usually 
restricted to wave functionals closely related to 
Gaussians, because the necessary path integrals can be done 
only in that case.  Also, this approach does not take advantage of the particle
spectrum that we expect to occur in quantum field theories.   For a 
relativistic theory, for example, the spectrum should consist of a
vacuum of energy-momentum zero, one or more single-particle states (including
possible bound states) with various masses (here, as in the rest of this paper,  
the analysis is restricted to cases without massless states), and many-particle 
states whose energies and momenta correspond to several massive particles.  
This particle structure can be put into a variational calculation at the outset 
by choosing to approximate the fields, rather than the states, and by using an
expansion in asymptotic fields for the interacting fields.  (Alternatively, as 
given below, one can use generalized free fields in the expansion of the 
interacting fields and {\it derive} the fact that the fields in the expansion 
are ordinary free fields.)  Thus, in quantum field theory, assuming 
completeness and irreducibility of the algebra of asymptotic fields, one finds
that the 
solutions lie in this algebra and the trial wave functions are replaced by 
trial operators chosen from this algebra.  Rudolf Haag \cite{haag} introduced 
the idea to use an element in the algebra of asymptotic fields to represent 
the interacting 
field.  It is fitting to call such a representation a ``Haag expansion.''  Some 
applications of Haag expansions are given in \cite{nqa}.  While a systematic
approximation using terms with normal-ordered products having arbitrarily
high degree in asymptotic fields leads to
amplitudes with arbitrarily many momenta and becomes intractable, a variational
trial operator in the algebra of asymptotic fields can have infinite degree, but
still can be parametrized in a tractable way.  Section 2 formulates the 
variational principle and proves that the minimum picks out the solution of the
field theory.  Section 3 illustrate the principle with the 
simplest cases: a free neutral scalar field of mass $m$ with 
the trial operator chosen to be a generalized free field, and a free Dirac field where 
the trial operator is a free Dirac field.  To compare with the variational method
using a vacuum functional, Sec. 4 studies the $\phi^4$ model in 1+1 and compares the 
results in the lowest approximation with the results of \cite{tik}. 
To show how the 
principle works in cases where the trial element has infinite degree, Sec. 5 applies 
the principle to the gradient coupling model, which has a ``nucleon'' Dirac 
field coupled to a scalar ``meson'' field.  In that case, one can choose the trial
operator for the nucleon field to be the product of an arbitrary function of a
free scalar field and a free nucleon field all at a single spacetime point and
the trial operator for the meson field to be a free scalar field.  Section 
concludes with the outlook for future work.

\section{The variational principle}

Using reasonable assumptions (in particular that no massless fields are 
present) and a physicist's level of mathematical rigor, one can show\cite{thesis} 
that the asymptotic fields obey the same Poincar\'e transformation law as the 
interacting fields.  For example, for a scalar field,
\begin{equation}
U(a, \Lambda)\phi^{as}(x)U(a, \Lambda)^{\dagger}=\phi^{as}(\Lambda x+a), \label{1}
\end{equation}
$\Lambda \in SO(1,3), ~ a \in R^4.$ For the general case,
\begin{equation}
U(a,A)\psi_i^{as}(x)U(a,A)^{\dagger}=
{\cal D}_{ij}(A^{-1})\psi_j^{as}(\Lambda x+a),
\end{equation}
where $A$ in $SL(2,C)$ is replaced by $\Lambda$ in $SO(3,1)$ 
for integer spin fields. 
One can also show that the asymptotic fields obey free field commutation relations,
for example, for a scalar field,
\begin{equation}
[\phi^{as}(x), \phi^{as}(y)]_-=i\Delta(x-y;m^2),                     \label{xcr}
\end{equation}
where
\begin{equation}
i\Delta(x,m^2)=(2 \pi)^{-3}\int d^4k \epsilon(k^0)\delta(k^2-m^2)
exp(-ik \cdot x)
\end{equation}
is the Pauli-Jordan commutator function.  These two 
conditions, together with the requirement that the vacuum have zero energy, 
imply that the generators of the Poincar\'e group are the free bilinear 
functionals of the asymptotic fields; in particular that 
(apart from a constant which
is the vacuum matrix element of the Hamiltonian) the Hamiltonian is 
diagonalized by the asymptotic fields,
\begin{equation}
P^0 \equiv H= const. + \sum_i H_{free}[\phi^{as}_i].                   \label{3}
\end{equation}
To see this, assume $H$ has an arbitrary expansion in asymptotic fields,
\begin{eqnarray}
H&=&F^{(0)}+\sum_{j=1}^nF_{-~j}^{(1)}A_j^{in}({\bf 0})+\sum_{j=1}^n
F_{+~j}^{(1)}A_j^{in}({\bf 0})^{\dagger}
\nonumber \\
& & +\sum_{s,t}[\int \frac{d^3k}{2 \omega_k}[F^{(2)}_{--~s,t}({\bf k})
A_s^{in}({\bf k})
A_t^{in}(-{\bf k})exp(-2i \omega_k x^0)+F^{(2)}_{+-~s,t}({\bf k})A_s^{in}({\bf
k})^{\dagger}A_t^{in}({\bf k}) \nonumber \\
 & &+F^{(2)}_{++~s,t}({\bf k})A_s^{in}({\bf k})^{\dagger}
A_t^{in}(-{\bf k})^{\dagger}exp(2i \omega_k x^0)] 
+\sum_{n>2}^{\infty}\int \frac{d^{3n}k}{\prod \omega_{k_i}} F^{(n)}(k_i) 
\nonumber \\
 & &\times :\prod_j A_j^{in}({\bf k}_i)^{()}:\delta(\sum \pm {\bf k}_i)
exp(i\pm \omega_{k_i}t \pm{\bf k}_i\cdot 
{\bf x}_i),~~\omega_k=\sqrt{{\bf k}^2+m^2}                 \label{h}
\end{eqnarray}
$A^{in}({\bf k}_i)^{()}$ stands for either the creation or the annihilation 
operator, normalized relativistically.  From Eq.(\ref{xcr}) and the Fourier 
transform
\begin{equation}
\phi^{in}(x)=(2 \pi)^{-3/2}\int d^4k  {\tilde \phi}^{in}(k)\delta_m(k) 
exp(-ik\cdot x)
\end{equation},
\begin{equation}
\phi^{(in)}(k)\delta_m(k)=\theta(k^0)\frac{A^{in}(\bf{k})}{2 \omega_{k}}
\delta(k^0-\omega_{k})+\theta(-k^0)\frac{A^{in}(-\bf{k})^{\dagger}}
{2 \omega_{k}}
\delta(k^0+\omega_{k}),
\end{equation}
the commutation relations of the creation and annihilation operators are
\begin{equation}
[A^{in}({\bf k}), A^{in~\dagger}({\bf l})]_-=
2\sqrt{{\bf k}^2+m^2}\delta({\bf k}-{\bf l})
\end{equation}                                                     \label{relcr}
The nonrelativistically normalized 
annihilation and creation operators are 
$a^{(in)}({\bf k}_i)^{()}=(2 E_k)^{-1/2}A^{in}({\bf k}_i)^{()}, E_k=k^0>0$. 
Put the form for $H$ into the infinitesimal form of Eq.(\ref{1}),
\begin{equation}
i[H, \phi^{as}(x)]_=\partial_0 \phi^{as}(x).                        \label{ascr}
\end{equation}
The commutation relation, Eq.(\ref{xcr} or \ref{relcr}), 
say that a normal-ordered term in $H$ with
$n$ factors of $\phi^{as}$ will contribute to a term with $n-1$ factors of 
$\phi^{as}$ on the right-hand-side of Eq.(\ref{ascr}).  Since the right-hand-side
of Eq.(\ref{ascr}) is linear in $\phi^{as}$,
the only terms allowed in $H$ are those with $n=0$ or $n=2$.  The 
$n=0$ term is the vacuum energy, and because $\phi^{as}$ is a free field, the 
$n=2$ term is the free field Hamiltonian.  

The Haag expansion of the interacting fields, stated for simplicity for a single
neutral scalar field with only scalar bound states, is
\begin{equation}
\phi(x)=\sum_{n=0}^{\infty} \frac{1}{n!} \sum_{j_i}
\int \sum_j d^{4n}x_i f^{(n)}(\{x-x_i\})
:\prod_{i=1} ^n \phi_{j_i}^{in} (x_i):,                                \label{hx}
\end{equation}
where the $\phi_{j_i}^{in}$ include in fields for stable bound states, if there 
are any.  The term {$f^{(0)}=v$} is a constant that occurs for a scalar field when 
symmetry is broken.  The term with {$f^{(1)}=1$} is just the in field with
coefficient one to fix field strength renormalization.  In momentum space, the 
expansion is
\begin{eqnarray}
\tilde{\phi}(k)&=&(2 \pi)^{3/2}v\delta(k)+\tilde{\phi}^{in}(k)\delta_{m}(k)  
+\sum^{\infty}_{n=2} \frac{1}{n!}\sum_j \int \tilde{f}^{(n)}(k_1,...,k_n)
\nonumber \\
& &\times :\prod_{i=1}^n\tilde{\phi}_j^{in}(k_i)\delta_{m_j}(k_i):
\delta(k-\sum k_i)\prod_{i=1}^n d^4k_i,                             \label{5}
\end{eqnarray}
$f^{(n)}(x_1,...,x_n)=(2 \pi)^{-3/2-5n/2}\int\tilde{f}^{(n)}(k_1,...,k_n)
exp(-i\sum k_i\cdot x_i)\prod d^4k_i$.  Hermiticity of the field implies
$\tilde{f}^{(n)}(k_1,...,k_n)=\tilde{f}^{(n) \ast}(-k_1,...,-k_n)$.  Similar
expansions hold for out fields.  When this expansion is inserted into the 
Hamiltonian, the result is an infinite series of the form already given in
Eq.(\ref{h}).
 
For a given Haag expansion, Eq.(\ref{5}), parametrized by $v$ and the 
${\tilde f}^{(n)}$'s, the $F^{(n)}$'s of Eq.(\ref{h}) are functionals of $v$ and
the ${\tilde f}^{(n)}$'s.  From the discussion above, for the 
exact solution $F^{(n)}=0$ for 
$n=1$ and for all $n>2$, $F^{(2)}=0$ for the $--$ and $++$ cases, and  
$F^{(2)}_{+-}$ is minimum.  In other words, as just discussed, 
the Hamiltonian is the sum of 
free-field Hamiltonians for each in field, together with a constant term which 
is the vacuum matrix element of $H$.  The condition that the Haag expansion
diagonalizes $H$ leads to an infinite set of nonlinear 
integral equations in $v$ and the ${\tilde f}^{(n)}$'s.  
A solution to this set of 
equations is equivalent to the solution of the field theory in all sectors.
In practice, it will be difficult to find an exact solution.  Variational methods
can yield approximate solutions.  Many different conditions can be imposed on the
$F^{(n)}$'s to find an approximation solution from a variational principle.
One such condition is to minimize 
the sum ${\cal Q}$ (with possible weighting factors $\lambda_n$)
of the integral of the absolute squares of the coefficients of the operator
terms, 
\begin{eqnarray}
{\cal Q}&= & \lambda_0|F^{(0)}|^2+\lambda_- |F_-^{(1)}|^2 + 
\lambda_+ |F_+^{(1)}|^2  \nonumber \\
& &
+\lambda_{--}\int d^3k |F^{(2)}_{--}(k,-k)|^2 + 
\lambda_{+-}\int d^3k|F^{(2)}_{+-}(k,k)|^2  \nonumber \\
& &+\lambda_{++}\int d^3k|F^{(2)}_{+-}(k,-k)|^2 +
\sum_n \lambda_n\int \prod d^3k_i|F^{(n)}(k_1,...,k_n)|^2     \label{q}
\end{eqnarray}
(here, to simplify the notation, I dropped the subscripts that label the
possibly different asymptotic fields).
The $\lambda$'s are arbitrary positive or vanishing 
numbers that can be chosen to control the weight attached to each term. 
If the Haag expansion of the interacting field has finite degree, the terms in
the Hamiltonian of highest degrees in in-fields cannot possibly vanish, so the
$\lambda$'s for such terms should be chosen to vanish.  For infinite degree
Haag expansions which remove the restriction to weak coupling one can keep all 
the $\lambda$'s positive.   For 
the exact solution all off-diagonal terms vanish and the diagonal terms are 
minimum, so this principle gives the exact solution of the theory {\it in all 
sectors} at the absolute minimum if all the $\lambda$'s are positive.  If all 
$\lambda$'s except the one multiplying the $a^{\dagger}a$ term for a bound 
state are chosen to vanish, this variational principle reduces to the quantum 
mechanical one for the bound state.  Another possible condition is to minimize 
the integral of the absolute square of the coefficients,
\begin{equation}
\int \prod d^3k_i|F^{(n)}(k_1,...,k_n)|^2,
\end{equation}
for some set of values of $n$ and for each combination of 
creation and annihilation parts of the
operators.  One can also choose to require that some set of sums of the 
positive terms in Eq.(\ref{q}) vanish.  Each of these possibilities 
gives a set of equations for
the solution of the variational principle.  The specific conditions one should
impose in a given problem should be chosen by experience.

\section{Application to the free field}

The simplest case on which to test this variational principle is the free field.
The Hamiltonian for a free neutral scalar field is
\begin{equation}\ 
H=\frac{1}{2} \int d^3x (\dot{\phi}^2 + (\nabla \phi)^2 + m^2 \phi^2)   \label{8}
\end{equation} 
As a trial operator, choose a generalized free field, $\phi_{gff}$,
whose two-point function has an unknown positive 
measure $\rho(\kappa^2)$.  To avoid the trivial case where $\rho=0$, require
$\int_0^{\infty} \rho(\kappa^2) d\kappa^2$ to have a fixed positive value; the 
exact value doesn't matter.  Let the field $\phi_{gff}$ be represented as the
sum of a term, $\phi_{dis}(x;\mu^2)$, with a {\it dis}crete weight at mass 
$\mu^2$ and
a term with a {\it con}tinuous weight $\sigma$, $\phi_{con}(x;\sigma)$, 
\begin{equation}
\phi_{gff}(x) = \phi_{dis}(x;\mu^2) + \phi_{con}(x;\sigma),
\end{equation}
\begin{equation}
\phi_{dis}(x;\mu^2) = \frac{1}{(2\pi)^{3/2}} \int \frac{d^3k}{2E_{\mu}({\bf k})}
 [A({\bf k}) e^{-ik \cdot x} + A^{\dagger}({\bf k})e^{ik \cdot x}],~~
k^0 = E_{\mu} ({\bf k}),
\end{equation}
\begin{eqnarray}
\phi_{con}(x;\sigma) & = & \frac{1}{(2\pi)^{3/2}} \int d \kappa^2
\frac{d^3k}{2E_{\kappa}({\bf k})} [B({\bf k}; \kappa^2)e^{-ik \cdot x}
+B^{\dagger}({\bf k}; \kappa^2)e^{ik \cdot x}],\nonumber \\
                     & k^0 & = E_{\kappa}({\bf k}) =   \sqrt{{\bf k}^2 + \kappa^2}.
\end{eqnarray}
The relativistically normalized commutation relations are
\begin{equation}
[A({\bf k}), A^\dagger({\bf \ell})]_-  =   2E_\mu({\bf k})
\delta({\bf k}-{\bf \ell}),                                     
\end{equation}
\begin{equation}
[B({\bf k};\kappa^2), B^\dagger({\bf \ell}; \lambda^2)]_- 
   =  2E_\kappa({\bf k}) \delta({\bf k}-{\bf \ell}) 
\delta(\kappa^2-\lambda^2)\sigma(\kappa^2),
\end{equation}
other commutators vanish.  To find the minimum of the energy of a given
particle, one should minimize the coefficient of the $a^{\dagger}a$ term.
The
relations between the relativistically (capital letters) and
nonrelativistically (lower-case letters) normalized operators are
\begin{equation}
A({\bf k}) =  \sqrt{2E_\mu({\bf k})}  a({\bf k}),
\end{equation}
\begin{equation}
 B({\bf k}; \kappa^2) =
 \sqrt{2E_\kappa({\bf k})}  b({\bf k}; \kappa^2).
\end{equation}
When the trial operator $\phi_{gff}$ is inserted into the free scalar
Hamiltonian, the result is
\begin{equation}
H = H_{dis} + H_{dis-con} + H_{con},
\end{equation}
\begin{eqnarray}
H_{dis} & = & \frac{1}{2} \int \frac{d^3k}{2E_{\mu}({\bf k})}
 (E_{\mu}({\bf k})^2 + {\bf k}^2+m^2) \delta({\bf 0}) \nonumber \\
& & + \frac{1}{2} \int \frac{d^3k}{2E_{\mu}({\bf k})}
 \{[({\bf k}^2+m^2 -E_{\mu}({\bf k})^2)
:a({\bf k})a(-{\bf k}): 
         e^{-2iE_{\mu}({\bf k})t}+ h.c. ]                    \nonumber \\
& &+2(E_{\mu}({\bf k})^2+{\bf k}^2 + m^2) :a^\dagger({\bf k}) a({\bf k}):\}
\end{eqnarray}
\begin{eqnarray}
H_{dis-con} & = & \int \frac{d\lambda^2d^3k}{2 \sqrt{E_\mu({\bf k}) 
E_\lambda({\bf k})} }  [(-E_\mu({\bf k})E_\lambda({\bf k}) 
          + {\bf k}^2 + m^2) :a({\bf k}) b(-{\bf k};\lambda^2): \nonumber \\
& & \times e^{-i(E_\mu({\bf k})+E_\lambda({\bf k}))x^0} 
            + (E_\mu({\bf k})E_\lambda({\bf k})+ {\bf k}^2 + m^2) \nonumber \\
& & \times :b^\dagger({\bf k};\lambda^2) a({\bf k}):
           e^{-i(E_\mu ({\bf k})-E_\lambda({\bf k}))x^0} +
            h.c.],
\end{eqnarray}
\begin{eqnarray}
H_{con} & = & \frac{1}{2} 
\int \frac{ d\kappa^2 d^3k}{2 E_{\kappa}{\bf k}}
(E_\kappa({\bf k})^2+\kappa^2+m^2) \sigma(\kappa^2) \delta({\bf 0})  
 + \frac{1}{2} \int \frac{d\kappa^2d\lambda^2d^3k}
  {2 \sqrt{E_\kappa({\bf k})E_\lambda({\bf k})} }  \nonumber \\
  &   & \times\{ [(-E_\kappa({\bf k})E_\lambda({\bf k})+ {\bf k}^2 +m^2)
:b({\bf k};\kappa^2) b(-{\bf k}; \lambda^2):    \nonumber\\
 &   &\times e^{-i(E_\kappa({\bf k}) +E_\lambda({\bf k}))x^0}+h.c.]
 +2(E_\kappa({\bf k})E_\lambda({\bf k}) + {\bf k}^2 +m^2)  \nonumber \\
  &   & \times:b^\dagger({\bf k};\lambda^2)b({\bf k};\kappa^2):
e^{-i(E_\kappa({\bf k}) -E_\lambda({\bf k}))x^0}\}
\end{eqnarray}
For the terms of the form 1 (vacuum energy) or $a^\dagger a$ (particle
energy), we must minimize the energy $(2{\bf k}^2+\mu^2+m^2)/\sqrt{{\bf k}^2+
\mu^2}$
 with respect to $\mu^2$, or, alternatively, minimize the same energy written as
$[E_{\mu}({\bf k})^2+ {\bf k}^2+m^2]/E_{\mu}({\bf k})$ with respect to 
$E_{\mu}({\bf k})$; for terms of the form $a a$ 
or $a^{\dagger} a^{\dagger}$ we minimize the absolute
value squared of the coefficient (since we want to bring
the Hamiltonian to diagonal
form).  For terms of the form $b^\dagger b$, we must
minimize the energy,
\begin{equation}
\frac{E_\kappa({\bf k})E_\lambda({\bf k})+{\bf k}^2+m^2}
{2 \sqrt{E_\kappa({\bf k})E_\lambda({\bf k})}},
\end{equation}
with respect to $\kappa^2$ and $\lambda^2$.
For terms of the form $b^\dagger a$ or $a^\dagger b$, similar expressions
have their minima for $\sigma$ concentrated at $\kappa^2 = \mu^2 = m^2$.
For terms of the forms $a a$, $a^\dagger a^\dagger$, $a b$, $a^\dagger
b^\dagger$, $b b$, and $b^\dagger b^\dagger$, the squares of the
coefficients vanish for $\mu^2 = m^2$ and $\sigma$ concentrated at
$\kappa^2 = m^2$.  For the contributions from $\phi_{con}$, the minimum
occurs for $\sigma$ at $\kappa^2 = m^2$.  The net result is
that the minimum of the operator H occurs at
\begin{equation}
\phi_{gff}(x)=\phi_{dis}(x;m^2), \phi_{con}=0;
\end{equation}  
i.e., for a free field of the mass in the Lagrangian.  This is no surprise.  In
general, for theories without massless fields or particles, we would assume free
field form for the asymptotic fields without doing a calculation.
(Note that these results require minimizing an expression that has
dimensions of energy.)

If one assumes a free field in the corresponding calculation of the Hamiltonian 
for
the Dirac field the $b^{\dagger}b$ and $d^{\dagger}d$ terms are diagonal in 
helicity; the $bd$ and $b^{\dagger}d^{\dagger}$ terms are not.  The minimum of 
the absolute valued squared coefficients also yields the result $\mu^2=m^2$.

\section{Application to the $\phi^4$ model in 1+1}  

A less trivial, but still elementary, example is the $\phi^4$ theory in one
space, one time dimension.  This model was studied recently using a variational
method by G. Tiktopoulos
\cite{tik}.  (The literature on variational calculations can be 
traced from \cite{tik,yee,kog,kk}.)
The example just below illustrates the variational principle in the 
algebra of asymptotic fields for this
theory in the lowest non-trivial approximation, where the calculations can be
done easily by hand.  Higher approximations require symbolic manipulation
programs that have been developed using Reduce 3.5.  We will 
report on the results of higher approximations separately.  The Lagrangian is
\begin{equation}
{\cal L}=\frac{1}{2}(\partial_{\mu}\phi\cdot\partial^{\mu}\phi-m_u^2 \phi^2)
- \frac{\lambda}{4!} \phi^4.
\end{equation}
The Hamiltonian density is 
\begin{equation}
{\cal H}=\frac{1}{2}(\dot{\phi}^2+ \phi^{\prime~2}+m_u^2\phi^2)+\frac{\lambda}{4!}\phi^4.
\end{equation}
For a theory that has no massless particles or fields, the
(on-shell) asymptotic fields will be an irreducible set of operators in which to 
expand the interacting field.  The demonstration given above that the free 
asymptotic fields diagonalize the free Hamiltonian supports this expectation.  The 
lowest variational ansatz is then
\begin{equation}
\phi(x)=v+\phi^{(in)}(x;m^2),
\end{equation}
where the constant $v$ is a symmetry-breaking vacuum matrix element of 
$\phi$ that
may occur due to radiatively-induced symmetry breaking and $\phi^{(in)}$ is 
a free 
field of unknown mass $m$.
 Substituting this
ansatz and re-normal ordering, the Hamiltonian becomes
\begin{eqnarray}
{\cal H}&=&[\frac{1}{2}(<\dot{\phi}_0^2>+<\phi_0^{\prime~ 2}>)+m_u^2
(v^2+<\phi_0^2>) \nonumber \\
& &+\frac{\lambda}{4!}(v^4+6v^2<\phi_0^2> +3<\phi_0^2>^2)] \nonumber \\
& &+[(m_u^2v+\frac{\lambda}{6}v^3)+\frac{\lambda}{2}v<\phi_0^2>]:\phi_0:
\nonumber \\
& &+\frac{1}{2}\{:\dot{\phi}_0^2+:\phi_0^{\prime~2}:+[m_u^2  
+\frac{\lambda}{2}(v^2+<\phi_0^2>)]:\phi_0^2:\}
\nonumber \\
& &+ \frac{\lambda}{6}v:\phi_0^3:+\frac{\lambda}{24}:\phi_0^4:,
\end{eqnarray}
where $\phi_0$ stands for $\phi^{(in)}(0;m^2)$.
The vacuum matrix element is just the first square bracket above.
The coefficient of the term in $(1/2):\phi _0^2:$ should be the square of the 
physical mass $m_v^2$, i.e.
\begin{equation}
m_v^2=m_u^2+\frac{\lambda}{2}(v^2+<\phi_0^2>).                      \label{gap}
\end{equation}
This is the gap equation in this approximation.
Following Tiktopoulos, one can verify this below by minimizing the vacuum 
energy density which is a functional of the energy, $E(k)$, of the trial in-field 
quanta of momenta $k$ with 
respect to this energy.  Regulate the vacuum and other divergent expressions with a
momentum-space cut-off; the two-point function is then
\begin{equation}
\langle \phi_0(x)\phi_0(y)\rangle=\frac{1}{2 \pi}\int_{-\Lambda}^{\Lambda}
\frac{dk}{2 E(k)}
e^{-iE(k)x^0+kx^1}.
\end{equation}
Then
\begin{equation}
 \langle\dot{\phi}_0^2\rangle=\frac{1}{4 \pi}\int_{-\Lambda}^{\Lambda}dk E(k),
~~ \langle\phi_0^{\prime~2}\rangle=\frac{1}{4 \pi}\int_{-\Lambda}^{\Lambda}dk 
\frac{k^2}{E(k)},~~
\langle\phi_0^2\rangle=\frac{1}{4 \pi}\int_{-\Lambda}^{\Lambda}dk \frac{dk}{E(k)}.
\end{equation}
 The vacuum energy density is
\begin{eqnarray}
<{\cal H}>&=&\frac{1}{8 \pi}\int_{-\Lambda}^{\Lambda}\frac{dk}{E(k)}
(E(k)^2+k^2+m_v^2)+\frac{1}{2}m_u^2v^2+\frac{\lambda}{24}v^4           \nonumber     \\
& &+\frac{\lambda}{16 \pi}v^2 \int_{-\Lambda}^{\Lambda}\frac{dk}{E(k)}
+\frac{\lambda}{2^7 \pi^2} (\int_{-\Lambda}^{\Lambda}\frac{dk}{E(k)})^2. 
\end{eqnarray}
The minimum is given by
\begin{eqnarray}
\frac{\delta \langle H \rangle}{\delta E(k)}&=&\frac{1}{8 \pi}
(1-\frac{k^2+m_u^2}{E(k)^2})
-\frac{\lambda}{16 \pi}\frac{v^2}{E(k)^2}-\frac{\lambda}{2^7 \pi^2}
\int_{-\Lambda}^{\Lambda}\frac{dk^{\prime}}{E(k^{\prime})}
\frac{1}{E(k)^2}           \nonumber   \\
&=& 0.
\end{eqnarray}
After multiplying by $8\pi E(k)^2$, the result is the gap equation, so 
$m^2=m_v^2$ as given by (\ref{gap}).

Since the gap equation relates the physical mass $m_v^2$, which should be
finite, to the bare mass and the divergent integrals cut-off at $\Lambda$, one can
follow Tiktopoulos in introducing another finite mass, $M$, via
\begin{equation}
M^2=m_u^2+\frac{\lambda}{8 \pi}\int_{-\Lambda}^{\Lambda}\frac{dk}{E(k)}.
\end{equation}
This form of renormalization can replace the usual one, since here only the
mass is renormalized.  As Tiktopoulos showed, the gap equation then relates
finite quantities,
\begin{equation}
m_v^2=M^2+\frac{\lambda}{2}v^2+\frac{\lambda}{8 \pi}ln\frac{M^2}{m_v^2}.
\end{equation}
The minimum of the vacuum energy density is
\begin{eqnarray}
\langle\cal{H}\rangle&=&\frac{1}{4 \pi}\int_{-\Lambda}^{\Lambda} dk |k|
-\frac{1}{2 \lambda}(M^2-\frac{\lambda}{8 \pi}
\int_{-\Lambda}^{\Lambda}\frac{dk}{\sqrt{k^2+m^2}})^2+\frac{1}{8 \pi} m_v^2   
\nonumber    \\
& & +\frac{\lambda}{24}v^4
+\frac{1}{2}v^2(M^2+\frac{\lambda}{8 \pi}ln\frac{M^2}{m_v^2})  
+\frac{1}{2\lambda}(M^2+\frac{\lambda}{8 \pi}ln\frac{M^2}{m_v^2})^2,
\end{eqnarray}
again in agreement with Tiktopoulos.  
Thus the present variational calculation leads
to the same results as does that of Tiktopoulos in the lowest approximation.
{ \it This agreement will not persist in higher approximations.}
Tiktopoulos adds Gaussians to his vacuum functional in order to improve
his approximation.  The Haag expansion suggests a different form of the 
higher-order terms, with higher-degree normal-ordered products of asymptotic
fields, in the expansion of the interacting field.  These are very different
approximations.  If the in (or out) fields are irreducible, then the Haag
expansion will approximate the exact solution in theories without massless
fields.  In higher approximations, the present variational method will place
emphasis on minimizing the absolute square of the coefficients of the 
non-quadratic terms in ${\cal H}$ and on minimizing the coefficients of
the $a^{\dagger} a$
terms which correspond to the energies of the asymptotic quanta.
  
\section{Application to the derivative coupling model in 1+3}

To illustrate a trial function that is of infinite degree, consider
the derivative coupling model, with
\begin{equation} 
{\cal L}=Z_2\bar{\psi}(i \not \!\partial + g \not \!\partial \phi - M)\psi
+\frac{1}{2}(\partial_{\mu}\phi \cdot \partial^{\mu} \phi - m^2\phi^2),
                                                                     \label{11}
\end{equation}
where $\phi$ and $\psi$ are renormalized fields.
The Hamiltonian is 
\begin{eqnarray}
H&=&\int d^3x[Z_2\bar{\psi}(i\gamma^j \partial^j +g\gamma^j \partial^j \phi+M)
\psi + \frac{1}{2}(\dot{\phi}^2+(\partial^j \phi)^2+m^2 \phi^2)]      \label{12}
\end{eqnarray}
Assume $\phi=\phi_0$, $\psi=:f(\phi_0):\psi_0$.  The Hamiltonian becomes
\begin{eqnarray}
H&=&\int d^3x[Z_2\bar{\psi_0}:f(\phi_0)^{\dagger}:(i:f^{\prime}(\phi_0)
\gamma^j\partial^j\phi_0:+
g\gamma^j \partial^j \phi_0 :f(\phi_0):+i:f(\phi_0):\gamma^j
\partial^j+ \nonumber\\
& &+M:f(\phi_0):)\psi_o  
 + \frac{1}{2}(\dot{\phi_0}^2+(\partial^j \phi_0)^2+m^2 \phi_0^2)].
                                                                      \label{13}
\end{eqnarray}
The first two terms in the bracket will cancel if
\begin{equation}
if^{\prime}+g f=0,                                                     \label{14}
\end{equation}
where the fact that $\langle \partial^j \phi \phi\rangle=0$ allows removing
$\partial^j \phi_0$ from the normal-ordered product in the first term.  
Field strength renormalization of $\psi$ requires $f(0)=1$.  The 
solution of this is {$f(x)=exp(igx).$}  
Evaluation of {$:f(\phi_0)^{\dagger}: :f(\phi_0):$} using the 
Baker-Hausdorff-Campbell theorem gives $exp(g^2 \langle \phi (x) \phi(y)\rangle)$ 
in the limit $x \rightarrow y$ for this
product .  This determines
\begin{equation} 
\frac{1}{Z_2}=
lim_{x \rightarrow y}exp(g^2 \langle \phi (x) \phi(y)\rangle).       \label{15}
\end{equation}
Thus the solution that reduces the 
Hamiltonian to free field form is 
\begin{equation}
\psi(x)=:exp(ig \phi_0(x)):\psi_0(x),~~\phi(x)=\phi_0(x).            \label{16}
\end{equation}
(Antisymmetrization of the Fermi fields $\psi$ and $\psi^{\dagger}$
above complicates
the calculation, but does not change the outcome.)

\section{Outlook for future work}

Much must be done to make this idea into a useful tool for field theory
calculations.  The method should be applied to examples with a finite number of
terms in the Haag expansion that serves as the trial element.  
Calculations of vertex functions in an all-scalar relativistic model
with the interaction $\phi^2 \chi$ are ongoing.  This method will be used to 
calculate hydrogen bound states with the soft photon cloud taken into account
for the charged particles.  The full power of the method will become
apparent only when infinite degree Haag expansions, parametrized in a tractable way,
are used.  The derivative coupling model discussed above is a trivial example of
such a parametrization. As mentioned above, the soft photon cloud 
around a charged particle will be represented using asymptotic fields in a later paper.
 Realistic models will be much more difficult to treat. 
We must also confront the problem of the coupling of high-energy and low-energy
modes pointed out by Feynman\cite{feyn}.  The step to nonabelian gauge theories
in which the confined fields do not have asymptotic fields will be the most
difficult step.  Hopefully the case of electrodynamics, where the charged fields
acquire a cloud of soft photons, which we believe we 
know how to treat using asymptotic fields, 
will serve as a stepping stone to the nonabelian case. 

{\bf Acknowledgements}

I am happy to thank Manoj Banerjee, Zacharia Chacko,  
Vigdor Teplitz,  
  and Ching-Hung Woo for helpful discussions.  I am greatly indebted
to George Tiktopoulos for extensive clarifications of his paper, as well as
for helpful comments about a draft of this paper.  I thank Shmuel Nussinov,
Daniel Phillips, and
Joseph Sucher for useful suggestions about an earlier version of this article.
It is a pleasure to thank Yasuo Umino for many discussions about this work, 
for collaboration in developing Reduce codes, as well as joint work on extensions
to other problems.


\begin{thebibliography}{99}

\bibitem{pes} M.E. Peskin and D.V. Schroeder, 
{\it An Introduction to Quantum Field
Theory} (Addison-Wesley, Reading, 1995).

\bibitem{wein} S. Weinberg, {\it The Quantum Theory of Fields}, Vol. I,
 Foundations;
Vol. II, Modern Applications (Cambridge, Cambridge, 1995, 1996).

\bibitem{fey} R.P. Feynman, Rev. Mod. Phys. {\bf 20}, 367 (1948); R.P. Feynman
and A.R. Hibbs, {\it Quantum Mechanics and Path Integrals}, 
(McGraw Hill, New York, 1965).

\bibitem{wil} K. Wilson, Phys. Rev. D {\bf 10}, 2445 (1974).

\bibitem{cre} M. Creutz, {\it Quarks, Gluons and Lattices}, 
(Cambridge, Cambridge, 1983).

\bibitem{tamm} I.E. Tamm, {\it J. Phys (USSR)} {\bf 9}, 449 (1945).

\bibitem{dan} S.M. Dancoff, {\it Phys. Rev.} {\bf 78}, 382 (1950).

\bibitem{bpp} S.J. Brodsky, H.-C. Pauli, and S.S. Pinsky, ``Quantum Chromodynamics
and Other Fields Theories on the Light Cone,'' hep-ph/9705477.

\bibitem{schiff} E. Merzbacher, {\it Quantum Mechanics}, (Wiley, New York, 1961).

\bibitem{tik} G. Tiktopoulos, ``Variational Wave Functionals in Quantum Field
Theory,'' hep-th/9705230.

\bibitem{yee} J.H. Yee, ``Variational Approach to Quantum Field
Theory: Gaussian Approximation and the Perturbative Expansion around It,''
hep-th/9707234.  

\bibitem{kog} W.E. Brown and I.I. Kogan, ``A Variational Approach to the QCD 
Wavefunctional: Calculation of the QCD $\beta$-Function,'' hep-th/9705136.

\bibitem{kk} I.I. Kogan and A. Kovner, Phys. Rev. D {\bf 51}, 1948 (1995) and
Phys. Rev. D {\bf 52}, 3719 (1995).

\bibitem{haag} R. Haag, {\it K. Dan Vidensk. Selsk. Mat-Fys. Medd.} {\bf 29},(12)
(1955).

\bibitem{nqa} O.W. Greenberg, {\it Phys. Rev. B} {\bf 139},1038 (1965); 
erratum, {\it Phys. Rev.} {\bf 156}, 1742 (1967); 
O.W. Greenberg and R. Genolio, {\it Phys.
Rev.}, {\bf 150}, 1070 (1966); A. Raychaudhuri, {\it Phys. Rev. D} {\bf 18}, 4658
(1978),  and University of Maryland thesis (1977); 
O.W. Greenberg, S. Nussinov and J.
Sucher, {\it Phys. Lett.} {\bf 70B}, 465 (1977); O.W. Greenberg,
{\it Prog. Theor. Phys. Supp.}  {\bf 86},  60 (1986); {\it Phys. Rev. D} 
{\bf 47},  331 (1993); O.W. Greenberg and P. K. 
Mohapatra, {\it Phys. Rev. D} {\bf 34}, 1136 (1986); O.W. Greenberg and L. Orr,
{\it Phys. Rev. D} {\bf 36}, 1240 (1987); O.W. Greenberg, R. Ray and F. 
Schlumpf, Phys. Lett. B {\bf 353}, 284 (1995). Related work appears in F. Gross, 
{\it Phys. Rev.} {\bf 186}, 1448 (1969); K. Johnson, {\it Phys. Rev. D} 
{\bf 4}, 1101 (1972); M. Bander, {\it Phys. Rev. Lett.} {\bf 47}, 549 (1981); 
{\bf 47}, 1419E (1981). 

\bibitem{thesis} O.W. Greenberg, {\it The Asymptotic Condition in Quantum
Field Theory}, Princeton Univ. PhD Thesis, 1956.

\bibitem{feyn} R.P. Feynman, in Proc. Int. Workshop on Variational Calculations
in Quantum Field Theory, (World Scientific, Singapore, 1988), ed. L. Polley and
D.E.L. Pottinger, p 28.

\end{thebibliography}
\end{document}